\documentclass[letterpaper, conference]{IEEEtran}
\usepackage{cite}
\usepackage{amsmath,amssymb,amsfonts}
\usepackage{algorithmic}
\usepackage{graphicx}
\usepackage{textcomp}
\usepackage{xcolor}
\usepackage{url}
\def\BibTeX{{\rm B\kern-.05em{\sc i\kern-.025em b}\kern-.08em
    T\kern-.1667em\lower.7ex\hbox{E}\kern-.125emX}}
\begin{document}

\title{A Machine Learning Framework for Large-Scale Static Wireless Mesh Networks
}

%\title{A Machine Learning Framework for Large-Scale Static Wireless Mesh Networks\\
%}

\author{\IEEEauthorblockN{Julia Andrusenko}
\IEEEauthorblockA{
\textit{Rampart Communications, Inc.}\\
Linthicum Heights, Maryland, United States \\
jandrusenko@rampartcommunications.com}
}

\maketitle

\begin{abstract}
This paper presents a system design methodology for a large-scale static wireless mesh network for 155 commercial off-the-shelf (COTS) radio nodes at fixed infrastructure sites in a challenging island environment. The architecture consists of approximately ten 15-node clusters, each with designated primary and secondary gateway nodes to support inter-cluster communication.

A structured, multi-stage planning methodology was developed to guide network design. Site-specific radio frequency (RF) path loss predictions were generated using Remcom’s Wireless InSite ray-tracing platform, incorporating terrain, buildings, and dense foliage effects. To optimize connectivity under physical-layer and operational constraints, spectral embedding combined with balanced k-means clustering was applied to partition the nodes into clusters of comparable size. A link budget analysis determined the maximum tolerable path loss under waveform and hardware constraints, defining the connectivity threshold used in the clustering framework.

This work integrates deterministic RF propagation modeling with constrained clustering optimization to provide a scalable framework for planning static wireless mesh networks in complex geographic environments. Node mobility and higher-layer networking protocols were outside the scope of this study.
\end{abstract}


\section{Introduction}
Large-scale, static wireless mesh networks provide resilient communications and monitoring for critical infrastructure in complex environments with challenging terrain, dense vegetation, and man-made structures. Such networks support situational awareness and operational continuity for systems including utilities and environmental monitoring. Maintaining reliable connectivity across distributed sites is difficult due to variable Radio Frequency (RF) propagation and physical-layer constraints, requiring a planning methodology grounded in predictive RF modeling and link analysis.

This paper presents the system design and initial planning framework for a wireless mesh network of 155 commercial off-the-shelf (COTS) radio nodes modeled as a representative fixed infrastructure network in a geographically complex island environment. To illustrate the methodology, we used 155 publicly available critical infrastructure locations, specifically Guam Water Authority (GWA) water well sites, sourced from the U.S. Geological Survey (USGS) \cite{usgs} and the Guam Hydrologic Survey (GHS) managed by the Water and Environmental Research Institute of the Western Pacific (WERI) \cite{ghs} as an open-source example for planning in complex propagation environment. Nodes are organized into ten clusters of approximately 15 nodes each, with each cluster containing designated primary and secondary gateway nodes to support inter-cluster connectivity.

The methodology integrates high-fidelity RF propagation modeling, connectivity-constrained clustering, and link budget analysis. We used the Remcom, Inc.'s Wireless InSite 3D ray-tracing model \cite{remcom_wi} to generate RF path loss predictions that account for terrain, structures, and foliage. Our prior studies have extensively validated this tool against measured wireless channel data. For example, when compared to actual measurements in scattering-rich urban environments, predicted path loss produced root mean square error (RMSE) values between 6.58 dB and 13.86 dB, which fall within the typical best-case accuracy range of approximately 12–15 dB RMSE reported for practical path loss models in the literature \cite{phillips2012}. In sparse desert environments, predictions showed cumulative mean errors within approximately 8 dB.

We then applied spectral embedding with balanced k-means clustering to partition nodes into size-controlled clusters based on predicted connectivity. A link budget analysis determined the maximum tolerable path loss, which established the connectivity threshold used during clustering.

The core contribution of this work is the development of a technical framework for the initial deployment and physical-layer planning of static wireless mesh networks in complex geographic areas. The framework is generalizable and provides a methodology for optimizing network resiliency and throughput. Node mobility and higher-level protocols, including routing algorithms, were beyond the scope of this initial study on the physical layer and node clustering.

\section{Relevant Work}
Wireless mesh networks (WMNs) enable scalable multi-hop wireless infrastructure. Prior studies have examined key design aspects, including gateway placement, connectivity, and capacity constraints in large-scale deployments \cite{akyildiz2005}, \cite{bicket2005}. Standards such as IEEE 802.11s define routing and frame structures for interoperable mesh networking \cite{IEEE2011}.

RF propagation modeling is essential for pre-deployment planning in environments with significant terrain and structural complexity. Deterministic ray-tracing techniques model reflections and diffractions to produce detailed, site-specific path loss estimates, though at higher computational cost than empirical or stochastic approaches. Their accuracy ultimately depends on the quality and completeness of the underlying input data, including geometry, materials, and environmental representation. Recent work has focused on optimizing ray-tracing methods to balance accuracy and runtime efficiency \cite{sandouno2023}.

Connectivity graphs abstract wireless links based on physical-layer feasibility. Spectral clustering methods have been applied to dense wireless networks to group nodes based on connectivity metrics rather than geographic proximity \cite{guedes2023spectral}. Such graph-based approaches provide a foundation for partitioning wireless systems into coherent clusters.

Classical clustering approaches in wireless systems include protocols such as LEACH \cite{heinzelman2000leach} and HEED \cite{younis2004heed}, which focus on energy efficiency and distributed operation in sensor networks rather than deterministic mesh pre-deployment design.

Multi-channel clustering in IEEE 802.11s networks explores adaptive cluster formation for interference mitigation and channel assignment under practical operating conditions \cite{rethfeldt2021chacha}. Mesh network planning approaches often rely on abstracted propagation and interference models to formulate optimization problems for coverage and connectivity \cite{amaldi2008wmnplanning}. Resilient mesh designs further address fault tolerance and interference management, often emphasizing protocol-level behavior rather than physical-layer modeling \cite{iqbal2014interference}.

In contrast, this work combines deterministic ray-tracing-based propagation modeling with spectral embedding and balanced k-means clustering to construct physically grounded partitions for static mesh deployments. We derive connectivity graphs from link-budget thresholds using Wireless InSite’s X3D shooting-and-bouncing-ray (SBR) model with GPU acceleration, multithreading, and exact path calculations. We enforce balanced cluster sizes and assign primary and secondary gateways to ensure resilient inter-cluster connectivity under link degradation or failure. To the best of our knowledge, few prior works jointly integrate high-fidelity propagation modeling, spectral clustering, and redundancy-aware mesh design for large-scale static deployments in complex environments.

\section{System Model and Link Budget Analysis}
\subsection{Propagation Modeling and Path Loss Data}
We used Wireless InSite software to generate site-specific path loss predictions for all 155 GWA node pairs at an operating frequency of 925 MHz, applying the X3D ray tracer with the Simple Attenuation foliage model and setting the specific attenuation to 0.18 dB/m based on ITU-R P.833-10 (Figure 2, vertical polarization) \cite{ITU833}.

The model incorporated detailed geographical inputs for Guam, including:
\begin{itemize}
    \item \textbf{\textit{Terrain:}} Wet Earth (conductivity = 2.00E-02 S/m, relative permittivity = 25)
    \item \textbf{\textit{Structures:}} Concrete buildings (conductivity = 0.015 S/m, relative permittivity = 7)
    \item \textbf{\textit{Foliage:}} Dense deciduous forest, in Leaf.
\end{itemize}
We used Wireless InSite software solely to generate unconstrained path loss data, without applying antenna gains or other hardware parameters during simulation. Hardware constraints were incorporated later in the link budget analysis. We modeled both transmitter and receiver antennas as omnidirectional, vertically polarized antennas installed at a 14-ft height with 0 dBi gain. This simulation produced the 155 × 155 path loss matrix used for subsequent clustering analysis.

\subsection{Link Budget and Maximum Tolerable Path Loss}
We performed a link budget analysis to constrain theoretical path loss using the physical characteristics of the considered radio hardware, focusing on a waveform of interest.

The hardware parameters used in the calculation included:
\begin{itemize}
    \item \textbf{\textit{Peak Output Power ($P_t$):}} 30.00 dBm
    \item \textbf{\textit{Tx/Rx Antenna Gains ($G_{t}, G_{r}$):}} 6.00 dBi
    \item \textbf{\textit{Total System Losses (cable, connector, and other losses):}} -11.62 dB.
\end{itemize}

Based on the required minimum receiver sensitivity for the selected waveform, we computed the maximum tolerable path loss ($PL_{max}$) corresponding to a 0 dB link margin as 143.79 dB. We used this $PL_{max}$ as the critical threshold to define usable wireless links in the network clustering methodology.

\section{Proposed Network Clustering Methodology}
The methodology for establishing the wireless mesh network relies on an unsupervised machine learning framework that transforms site-specific path loss data into a measure of network connectivity, enabling the formation of balanced sub-meshes (clusters) and the designation of inter-cluster gateway nodes.

We used Wireless InSite software to synthesize path loss, MATLAB \cite{matlab} to construct the similarity matrix (noting this step could also be implemented in Python), and Python libraries including \texttt{scikit-learn}~\cite{scikit-learn} and NumPy~\cite{numpy} to implement spectral embedding with balanced k-means clustering.

\subsection{Data Preparation and Symmetric Path Loss Matrix}
The process began with the creation of a symmetric $155 \times 155$ path loss matrix ($PL$) for these nodes. The original path loss data generated by Wireless InSite software exhibited slight non-reciprocity between node pairs (Node A $\rightarrow$ Node B $\neq$ Node B $\rightarrow$ Node A). To ensure symmetry for the subsequent clustering algorithms, we populated the matrix using the maximum path loss value for each reciprocal radio pair. We also set all non-zero self-pair entries (i.e., the diagonal) to zero. The requirement for a symmetric path loss matrix ($PL$) ensures consistent pairwise link representation, which is essential for constructing a valid similarity matrix ($S$). We use this similarity matrix as the weighted adjacency matrix of the graph, which then serves as input to spectral embedding and balanced k-means clustering for network partitioning.

\subsection{Similarity Matrix Construction}

The path loss data, a measure of dissimilarity, was converted into a similarity matrix ($S$). This matrix acts as a weighted graph adjacency (affinity) matrix where each node represents a radio, and the edge weight represents the strength of the connection between nodes. Stronger connections (low path loss) result in high similarity scores, while weak connections (high path loss) result in scores approaching zero.

This transformation uses the exponential decay equation:

\begin{equation}
S_{ij} = e^{-\alpha \, PL_{ij}}
\end{equation}

where \(S_{ij}\) is the similarity between nodes \(i\) and \(j\), \(PL_{ij}\) is the path loss between nodes \(i\) and \(j\) (\(PL \in \mathbb{R}^{155 \times 155}\)), and \(\alpha\) is a scaling parameter that controls the rate of similarity decay. This exponential mapping is consistent with commonly used kernel-based similarity constructions in spectral clustering, where dissimilarity measures are converted into affinities using decaying exponential functions \cite{belkin2001laplacian, favati2020}.

The scaling parameter $\alpha$ was chosen to ensure that the exponential similarity spans the dynamic range of the path loss data based on hardware limitations, determined by the maximum tolerable path loss ($PL_{\text{max}} = 143.79\,\text{dB}$) for our waveform of interest. $\alpha$ is computed as:

\begin{equation}
\alpha = \frac{\ln(\text{sim}_{\text{lo}} / \text{sim}_{\text{hi}})}{PL_{\text{max}} - PL_{\text{min}}}
\end{equation}

where $\text{sim}_{\text{lo}} = 0.9$ is the desired similarity for the lowest path loss, $\text{sim}_{\text{hi}} = 0.01$ is the desired similarity for the worst usable path loss, and $PL_{\text{min}}$ is the minimum usable path loss ($> 0\,\text{dB}$). We excluded self-similarity terms by construction, i.e., we set $S_{ii} = 0$, since path loss between a node and itself had no physical meaning in the propagation model. As a result, the similarity matrix captured only pairwise interactions between distinct nodes, and we did not include self-loop contributions in subsequent spectral processing.

\subsection{Node Clustering Methodology}

As stated previously, we performed node clustering using spectral embedding followed by a balanced k-means algorithm. This approach addressed a limitation in standard spectral clustering, which did not reliably produce equal-sized clusters of approximately 15 nodes in our case study. We implemented the pipeline in Python, using spectral embedding and standard k-means from \texttt{scikit-learn}, while implementing our own balanced k-means algorithm using NumPy.

Spectral embedding in \texttt{scikit-learn} constructs a graph Laplacian from the similarity matrix $S$ and, by default, uses the symmetric normalized form described in \cite{williamson2016lecture7}:
\begin{equation}
\mathcal{L}_{sym} = I - D^{-1/2} S D^{-1/2}.
\end{equation}

The degree matrix $D$ is computed internally within the spectral embedding routine as $D_{ii} = \sum_j S_{ij}$. The \texttt{SpectralEmbedding} function then implicitly forms and subsequently eigendecomposes this Laplacian. The embedding is obtained from the eigenvectors associated with the smallest non-zero eigenvalues, yielding a low-dimensional representation of the graph.

This representation preserves the underlying RF connectivity structure by mapping strongly coupled nodes to nearby points in the embedding space, while separating weakly connected nodes.

We then performed clustering in the embedded space using a balanced k-means algorithm implemented in Python. We initialized cluster centroids using k-means from \texttt{scikit-learn}, which minimizes within-cluster Euclidean distance. Standard k-means does not enforce equal cluster sizes.

To address this, we applied a capacity-constrained assignment step consistent with balanced k-means formulations, where each cluster was limited to approximately 15 nodes. Each point was assigned to the nearest centroid subject to these capacity constraints.

This method preserves the geometry induced by the spectral embedding while enforcing balanced cluster sizes, enabling a uniform partitioning of the RF graph for gateway selection and network organization.

\subsection{Gateway Node Selection and Inter-Cluster Connectivity}

To enable inter-cluster communication, we identify a primary and a secondary gateway node within each cluster based on their aggregate connectivity to nodes in other clusters. These gateway nodes serve as bridge points between the sub-meshes formed by the balanced spectral clustering procedure.

To quantify external connectivity, we defined a gateway score for each node $i$ as the sum of its similarity to all nodes outside its assigned cluster. The gateway score is computed as:

\begin{equation} \label{eq:gateway_score}
\text{GatewayScore}_i = \sum_{j \notin \mathcal{C}(i)} S_{ij},
\end{equation}

where $\mathcal{C}(i)$ denotes the cluster assignment of node $i$, and $S_{ij}$ represents the similarity between nodes $i$ and $j$. This metric captures how strongly a node is connected to nodes in other clusters induced by the balanced spectral embedding and assignment procedure.

Within each cluster, nodes are ranked according to their gateway scores. The node with the highest score is selected as the primary gateway, and the node with the second-highest score is selected as the secondary gateway when available.

We also defined a reduced similarity matrix by restricting the original similarity matrix to the gateway set $\mathcal{G}$, which contains all primary and secondary gateway nodes across clusters. This gateway-level matrix captures pairwise similarities between gateway nodes and enables analysis of connectivity structure at a reduced level.

We construct the gateway-restricted similarity representation by extracting the corresponding submatrix:

\begin{equation}
\mathbf{S}_{\mathcal{G}} = \mathbf{S}[\mathcal{G}, \mathcal{G}],
\end{equation}

where $\mathbf{S}_{\mathcal{G}}$ preserves pairwise similarities among gateway nodes and retains the geometric structure induced by the spectral embedding of the full graph.

For analysis of inter-cluster interactions only, we defined a filtered similarity matrix by removing intra-cluster contributions:

\begin{equation}
S^{\text{inter}}_{ij} =
\begin{cases}
S_{ij}, & C(i) \neq C(j) \\
0, & C(i) = C(j)
\end{cases}
\end{equation}

We then obtained the inter-cluster gateway representation by restricting the filtered similarity matrix to the gateway set:

\begin{equation}
\mathbf{S}^{\text{inter}}_{\mathcal{G}} =
\mathbf{S}^{\text{inter}}[\mathcal{G}, \mathcal{G}],
\end{equation}

which captures gateway interactions across clusters in a reduced form.

\section{Results and Discussion}

\begin{figure}[htbp]
\centering
\includegraphics[width=\columnwidth]{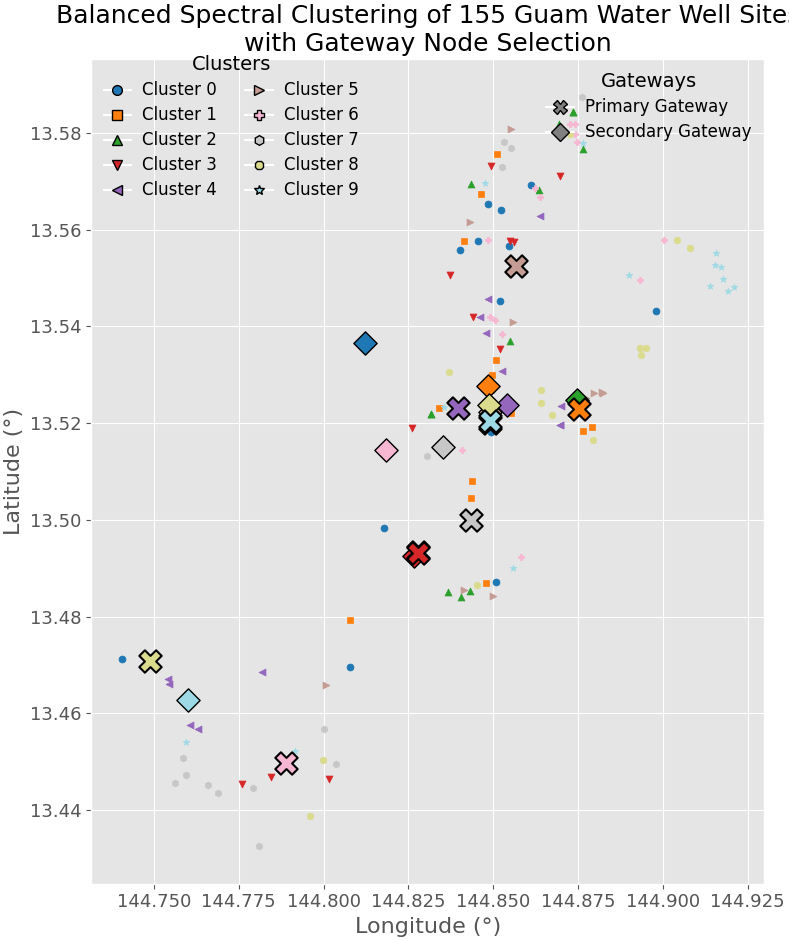}
\caption{Guam Radio Node Clustering with Primary and Secondary Gateways.}
\label{fig1}
\end{figure}

We successfully partitioned the 155 GWA wells into 10 sub-meshes (i.e., clusters) by enforcing balanced cluster sizes, ensuring a uniform partitioning of the RF graph. Fig.~\ref{fig1} depicts the resulting geographic partitioning and cluster assignments of the GWA nodes, along with the designated gateway nodes.

\begin{figure}[htbp]
\centering
\includegraphics[width=\columnwidth]{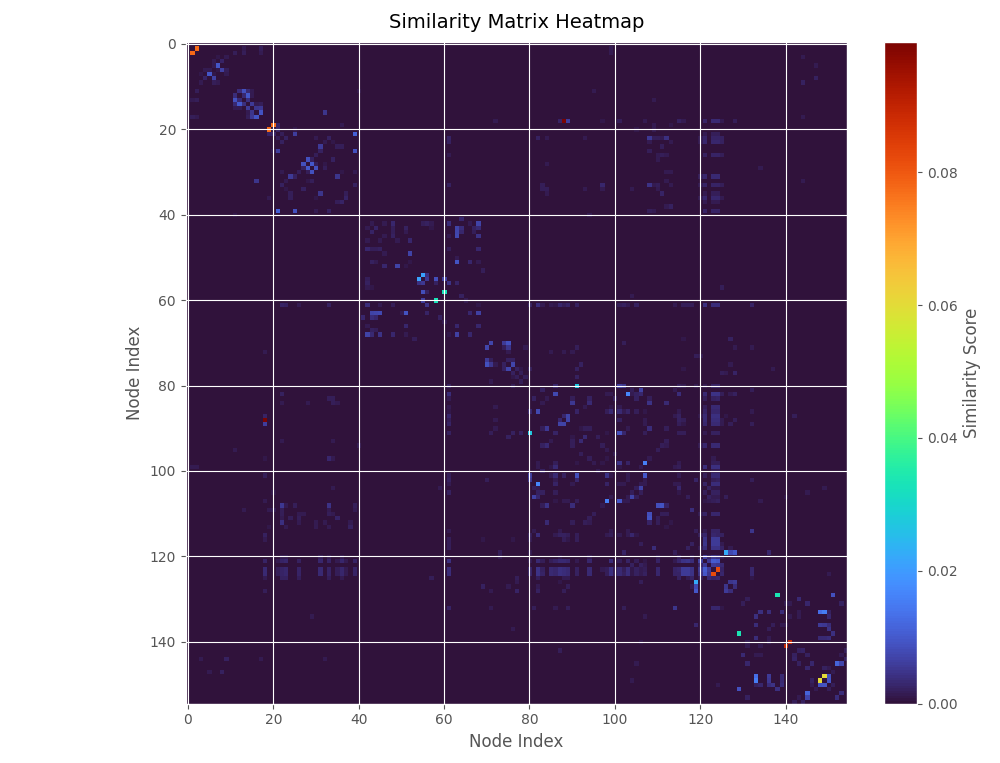}
\caption{Full Similarity Matrix.}
\label{fig2}
\end{figure}

Fig.~\ref{fig2} shows the full similarity matrix ($S$) heatmap, which exhibits uniformly low similarity scores across node pairs, consistent with high path loss between the GWA wells. These low similarity scores reflect the challenging RF environment in Guam, including complex topography, buildings, and dense jungle foliage, all of which were incorporated into the Wireless InSite propagation model.

\begin{figure}[htbp]
\centering
\includegraphics[width=\columnwidth]{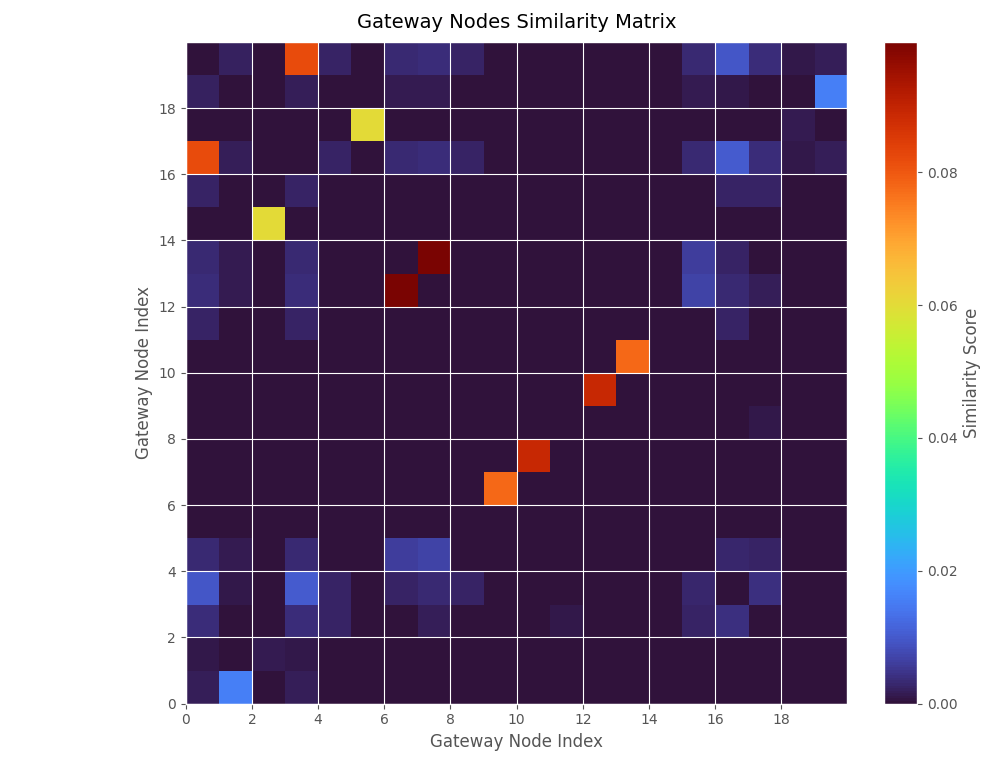}
\caption{Gateway-Restricted Similarity Matrix.}
\label{fig3}
\end{figure}

Lastly, Fig.~\ref{fig3} illustrates Guam’s gateway-restricted similarity matrix $\mathbf{S}_{\mathcal{G}}$, which, as a subset of the full similarity matrix $S$, also exhibits low similarity values due to the challenging island geometry and propagation environment.

\section{Conclusion}
This paper presented a preliminary system design methodology for a static radio node network, using the 155 water well sites in Guam as a case study. We established a foundation for a wireless mesh network comprising approximately ten 15-node sub-meshes with designated primary and secondary gateway nodes.

The proposed framework integrated high-fidelity RF propagation modeling (Wireless InSite software) with an unsupervised machine learning pipeline, combining spectral embedding and balanced k-means clustering to achieve balanced partitioning. A link budget analysis for the waveform parameters under study determined a maximum tolerable path loss of $143.79$ dB at a 0 dB link margin, which directly informed construction of the connectivity graph used for clustering.

The methodology successfully identified primary and secondary gateway nodes, providing a structural basis for resilient inter-mesh communication. Higher-layer networking functions, including routing protocol design, were outside the scope of this study. Future work will extend this framework to incorporate node mobility and develop and evaluate mesh routing protocols. Overall, the framework is generalizable and provides an initial blueprint for physical-layer, geography-aware network design in complex environments and static deployments.

\end{document}